\documentclass[12pt]{article}
\pdfoutput=1

\usepackage{amsmath,amssymb,graphicx} 
\usepackage{epsf}
\usepackage{pstricks}
\usepackage{cite}
\usepackage{multirow}
\usepackage{array}

\newcommand{\beq}{\begin{eqnarray}}
\newcommand{\eeq}{\end{eqnarray}}
\definecolor{orange}{rgb}{1, 0.4, 0} 
\definecolor{vertfonce}{rgb}{0, 0.4, 0} 
\definecolor{marron}{rgb}{0.36,0.13,0.00} 
\definecolor{purple}{rgb}{0.4,0.0,0.4} 
\definecolor{pink}{rgb}{0.8,0.3,0.6} 
\definecolor{gray}{rgb}{0.3,0.3,0.3}

\begin{document}
\begin{titlepage}
\begin{flushright}
LYCEN 2012-11 
\end{flushright}

\vskip.5cm
\begin{center}
{\huge \bf 
Higgs couplings beyond the Standard Model}

\vskip.1cm
\end{center}
\vskip1cm

\begin{center}
{\bf
{Giacomo Cacciapaglia}, {Aldo Deandrea}, {Guillaume Drieu La Rochelle} and {Jean-Baptiste Flament}}
\end{center}
\vskip 8pt

\begin{center}
{\it
Universit\'e de Lyon, F-69622 Lyon, France; Universit\'e Lyon 1, Villeurbanne;\\
CNRS/IN2P3, UMR5822, Institut de Physique Nucl\'eaire de Lyon\\
F-69622 Villeurbanne Cedex, France} \\
\vspace*{.5cm}
{\tt  g.cacciapaglia@ipnl.in2p3.fr, deandrea@ipnl.in2p3.fr, g.drieu-la-rochelle@ipnl.in2p3.fr, j-b.flament@ipnl.in2p3.fr}
\end{center}

\vskip2truecm

\begin{abstract}
\vskip 3pt
\noindent
We consider the Higgs boson decay processes and its production, and provide a parameterisation tailored for testing models of new 
physics beyond the Standard Model. We also compare our formalism to other existing parameterisations based on scaling factors in 
front of the couplings and to effective Lagrangian approaches. Different formalisms allow to best
address different aspects of the Higgs 
boson physics. The choice of a particular parameterisation depends on a non-obvious balance of quantity and quality of the available 
experimental data, envisaged purpose for the parameterisation and degree of model independence, importance of the radiative 
corrections, scale at which new particles appear explicitly in the physical spectrum. At present
only simple parameterisations with a 
limited number of fit parameters can be performed, but this situation will improve with the forthcoming experimental LHC data. 
Detailed fits can only be performed by the experimental collaborations at present, as the full information on the different decay modes 
is not completely available in the public domain. It is therefore important that different approaches are considered and that the most 
detailed information is made available to allow testing the different aspects of the Higgs boson physics and the possible hints beyond the Standard Model.

\end{abstract}

\end{titlepage}

\newpage

\section{Higgs coupling parameterisations}

The discovery of a new resonance at a mass of $125$ GeV, announced by both ATLAS and CMS in July this year, has opened a new era in particle physics. 
In fact, the new particle has similar behaviour as the one expected for a Standard Model (SM) Higgs
boson, and it has been observed in different channels, notably the decay into a pair of photons and
into a pair of massive gauge bosons ($ZZ$ and $W^+ W^-$, where one of the two vectors is virtual).
If confirmed, the discovery of the hard-sought SM Higgs boson would complete the picture for the Standard Model.
However, intriguing discrepancies have also been observed, like for instance an excess in the di-photon rate and the non-observation of di-tau signal events by CMS.
Such discrepancies are not statistically significant, thus they may disappear as mere statistical
fluctuations.
Even in this case, knowing the couplings of the new resonance is a crucial test for the SM hypothesis.
Furthermore, many models of new physics, especially the ones addressing the problem of the hierarchy in the electroweak symmetry breaking sector, predict sizeable deviations in the Higgs couplings.
Precise measurements of the Higgs properties, therefore, can give precious information on the kind of new physics that Nature chose.
The information that can be extracted at the LHC is rather limited to a few channels, nevertheless it is important to make the best out of it.
In the light of this consideration, it is important to choose the most relevant parameterisations for the deviations from the SM couplings, and use them to present the LHC measurements.
Higgs coupling parameterisations, meant to perform fits on the available data, may follow different
approaches which are not 
completely independent. 
In the following we will discuss a few of them without aiming at an exhaustive description. We shall then discuss our suggestion which is particularly motivated by testing models Beyond the Standard Model (BSM). 

From the experimental point of view it makes sense to just parameterise Higgs physics in terms of observed quantities such as 
branching ratios and cross-sections. This is for example the case of the parameterisation proposed in 
Ref.~\cite{LHCHiggsCrossSectionWorkingGroup:2012nn}, where the relevant cross-sections and partial decay widths are multiplied by a suitable factor. The advantage of such an approach is its simple link to the experimentally 
measured quantities. On the other hand, with such a choice, correlations among the different parameters are not explicit, in particular between tree level and loop induced observables. For 
example, a modification of the couplings to $W$ bosons and top, while modifying tree-level branching ratios and cross-sections for the Higgs boson, can also affect the loop-level couplings for the Higgs production via the 
gluon channel or the Higgs decay into two photons. 
This is in principle not a limitation, but in practice if one wants to take these 
correlations into account a different choice of parameters needs to be considered.
Another point is: what if, instead of a generic fit, one aims at discussing limits for a particular BSM model? In that case explicit 
correlations among the fit parameters should be calculated in order to correctly compute the number of independent degrees of 
freedom for the fit.
Therefore, a parameterisation which can easily be connected to any model of new physics is also useful.
Along these lines, we propose an extension of the parameterisation in
Ref.~\cite{Cacciapaglia:2009ky}, where the contribution of loops of New Physics to the $H \to g g$
and $H \to \gamma \gamma$ modes is explicitly disentangled from the modification of tree level
couplings, thus removing correlations among the various parameters.
Furthermore, the loop contributions are normalised to the top ones, thus simplifying the interpretation in terms of new models.

Another point of view consists in parameterising physics in terms of effective operators (for a specific strategy concerning the Higgs 
boson data and the interpretation in terms of physics beyond the Standard Model 
see \cite{Bonnet:2011yx,Bonnet:2012nm, Corbett:2012dm}).  Another example is 
given by chiral electroweak Lagrangians \cite{Espinosa:2012ir,Espinosa:2012im}. This approach has the advantage of inheriting all the standard know-how in effective theories, 
including the calculation of radiative corrections. The possible remarks to such an approach is the large number of effective 
couplings when going beyond the lowest order set and the treatment of possible light degrees of freedom beyond the SM particles 
(this possibility is not completely ruled out as particles in the same mass range as the SM Higgs boson might still be possible, see for 
example \cite{Belanger:2012tt}). A quite detailed and clearly written overview of effective Lagrangians for Higgs physics is given in 
\cite{Passarino:2012cb}, including the treatment of radiative corrections.

The question of the choice of a parameterisation does not only depend on the particular use or preference for a given formalism, 
but also depends crucially on the number of fit parameters with respect to the number of physically independent data channels. 
In the frequentist approach one computes a $\chi^2$ per degree of freedom within a given choice of model (with a given number 
of independent fit parameters). The choice of a particular fit ``model'' is a matter of a reasonable rule of thumb, as too many fit 
parameters compared to too few experimental data channels give a poor quality fit \footnote{``With four parameters I can fit an elephant, 
and with five I can make him wiggle his trunk''.  Rule of thumb attributed to John von Neumann by Enrico Fermi, as quoted by 
Freeman Dyson in ``A meeting with Enrico Fermi'' in Nature 427 (January 2004) p. 297.}. 

In the following we wish to discuss how the parameterisation of the Higgs couplings to gluons and photons proposed in 
Ref.~\cite{Cacciapaglia:2009ky} can be extended to include tree level couplings modifications, and how it compares to 
other parameterisations, in particular the one proposed in Ref.~\cite{LHCHiggsCrossSectionWorkingGroup:2012nn}. A similar study has also been recently carried out in \cite{Belanger:2012gc}. 
We will try to show that the extension of \cite{Cacciapaglia:2009ky} is especially useful when the results of a fit on the parameters 
are to be interpreted in terms of specific models of new physics beyond the standard model.
For the tree level couplings, we follow the same parameterisation as in Table~2 of 
Ref.~\cite{LHCHiggsCrossSectionWorkingGroup:2012nn}, i.e. we introduce a scaling factor in front of the coupling, 
$\kappa_X$ where $X$ is any massive particle of the SM the Higgs couples directly to.
The same scaling factor will appear in front of some cross sections and partial decay widths.
For example:
\beq
\sigma_{Wh} = \kappa_W^2 \sigma_{Wh}^{SM}\,, \quad
\sigma_{Zh} = \kappa_Z^2 \sigma_{Zh}^{SM}\,, \quad
\sigma_{t\bar{t}h} = \kappa_t^2 \sigma_{t\bar{t}h}^{SM}\,.
\eeq
For the partial decay widths:
\beq
\Gamma_{WW} = \kappa_W^2 \Gamma_{WW}^{SM}\,, \quad
\Gamma_{ZZ} = \kappa_Z^2 \Gamma_{ZZ}^{SM}\,, \quad
\Gamma_{b\bar{b}} = \kappa_b^2 \Gamma_{b\bar{b}}^{SM}\,, \quad
\Gamma_{\tau^+ \tau^-} = \kappa_\tau^2 \Gamma_{\tau^+ \tau^-}^{SM}\,, \dots
\eeq
For the Vector Boson Fusion (VBF) cross sections, it is imperative to distinguish the two production channels with $W$ or $Z$ fusion:
\beq
\sigma_{VBF} = \kappa_W^2 \sigma_{WF}^{SM} + \kappa_Z^2 \sigma_{ZF}^{SM}\,.
\eeq
So far, the parameterisation is the same as in Ref.~\cite{LHCHiggsCrossSectionWorkingGroup:2012nn}: the crucial differences 
arise in the treatment of loop induced couplings, as explained in the following Section. 

\section{Loop induced couplings: $\kappa_{gg}$ and $\kappa_{\gamma\gamma}$ vs. $\kappa_g$ and $\kappa_\gamma$}

The parameters introduced so far describe tree level couplings of the Higgs.
Typically, sizeable modification to such couplings are generated by tree level effects from New Physics, like for example mixing of the SM particles with heavier states.
Modifications of the loop induced couplings, however, deserve a different treatment, because they are directly sensitive to any new state that may enter the loop.
In Ref.~\cite{LHCHiggsCrossSectionWorkingGroup:2012nn}, a scaling parameter was also introduced to describe the new physics 
effects in the couplings to gluons and photons, namely:
\beq
\sigma_{ggH} = \kappa_g^2 \sigma_{ggH}^{SM}\,, \quad \Gamma_{gg} = \kappa_g^2 \Gamma_{gg}^{SM}\,, \quad \Gamma_{\gamma \gamma} = \kappa_\gamma^2 \Gamma_{\gamma \gamma}^{SM}\,.
\eeq
However, both $\kappa_g$ and $\kappa_\gamma$ depend non trivially on the tree level couplings, in particular $\kappa_W$ and 
$\kappa_t$, because a modification of the couplings to $W$ and tops would affect the SM loop contribution to the couplings to gluons and photons. So, in general, there is a correlation intrinsic in this scaling parameter approach.

On the contrary, in Ref.~\cite{Cacciapaglia:2009ky}, we proposed an alternative parameterisation of the couplings to gluons and photons that can deal with 
loop corrections from new physics in an independent way from the tree level corrections to the couplings to massive SM states.
In this way, the parameters are not correlated to each other. Note that a parameterisation equivalent to the one we present here has 
been recently and independently used in \cite{Moreau:2012da}.
Furthermore, our parameterisation allows to obtain bounds that are more easily interpreted in terms of new physics models.
The new parameters, $\kappa_{gg}$ and $\kappa_{\gamma \gamma}$, enter at the level of the amplitude of the loop corrections.
In terms of the partial decay widths, we have:
\beq
\Gamma_{\gamma \gamma} &=& \frac{G_F \alpha^2 m_H^3}{128 \sqrt{2} \pi^3} \left|  \kappa_W \, A_W (\tau_W) + 
C^\gamma_t \; 3 \left( \frac{2}{3} \right)^2 A_t (\tau_t)\; [\kappa_t+\kappa_{\gamma \gamma} ] + \dots \right|^2\,, \\
\Gamma_{g g} &=& \frac{G_F \alpha_s^2 m_H^3}{16 \sqrt{2} \pi^3} \left| C^g_t \frac{1}{2} A_t (\tau_t)\; [\kappa_t+\kappa_{gg}] + \dots \right|^2\,,
\eeq
where the dots stand for the negligible contribution of the light quarks. The coefficients $C^\gamma_{t}$ and $C^g_t$ contain the NLO 
QCD corrections to the SM amplitudes. $A_W$ and $A_t$ are the well known $W$ and top amplitudes:
\beq
A_t (\tau) &=& \frac{2}{\tau^2} \left( \tau + (\tau-1) f (\tau) \right)\,, \\
A_W (\tau) &=& - \frac{1}{\tau^2} \left( 2 \tau^2 + 3 \tau + 3 (2 \tau - 1) f (\tau) \right)\,, 
\eeq
where $\tau = \frac{m_H^2}{4 m^2}$ and
\beq
f (\tau) = \left\{ \begin{array}{lc}
\mbox{arcsin}^2 \sqrt{\tau}  & \tau \leq 1 \\
- \frac{1}{4} \left[ \log \frac{1+\sqrt{1-\tau^{-1}}}{1-\sqrt{1-\tau^{-1}}} - i \pi \right]^2 & \tau > 1 \end{array} \right.\,.
\eeq
The amplitudes $A_W$ and $A_t$ have the property that they rapidly asymptotise to a constant value for large masses of the states inside the loop, i.e. for small $\tau$.
For a Higgs mass of $125$ GeV, we find:
\beq
A_W (\tau_W) = -8.32\,, \qquad A_t (\tau_t) = 1.37\,;
\eeq
where the top amplitude is very close to its asymptotic value $A_t (0) = 4/3 \sim 1.33$.
The $W$ amplitude however significantly deviates from the asymptotic value $A_W (0) = - 7$.
The QCD corrections to the top loops can also be computed in the asymptotic limit of large top mass, which gives a good approximation of the much more complicated mass dependent corrections.
One finds~\cite{Spira:1995rr}:
\beq
C^\gamma_t = 1- \frac{\alpha_s}{\pi} \,, \qquad C^g_t = 1+\frac{9}{2} \frac{\alpha_s}{\pi} \,.
\eeq
In the case of the gluon loop, the corrections, which also includes real emission of an additional gluon and splitting into a pair of light quarks, reduces to this simple form if $\alpha_s$ entering the LO amplitude is evaluated at the renormalisation scale $\mu = e^{-7/4} m_H \sim 22$ GeV: in other words, part of the correction is encoded in the running of the coupling constant.
A more detailed discussion of the QCD corrections that can be included in this parameterisation will be discussed in a following Section.

The original simplified parameterisation of \cite{Cacciapaglia:2009ky} can be recovered setting the NLO coefficients 
$C^\gamma_{t}$ and $C^g_t$ to one and $\kappa_W=1$, $\kappa_t=1$. Note that the contribution of $\kappa_t$ and 
$\kappa_W$ was effectively included into the loop parameters, and we will discuss in more detail later in which case this procedure is allowed.
We have normalised the contribution of the new physics loops to the contribution of the top loop alone.
This is important when we want to interpret the fit of the parameters in terms of the properties and nature of the new physics running into the loop.

Neglecting the contribution of light fermions, we can draw a relation between the 
parameterisation \cite{LHCHiggsCrossSectionWorkingGroup:2012nn} and ours:
\beq
\kappa_g (\kappa_t, \kappa_{gg}) &=& \left| \kappa_t+\kappa_{gg} \right|\,, \\
\kappa_\gamma (\kappa_W, \kappa_t, \kappa_{\gamma \gamma}) &=& 
\left| \frac{ \kappa_W \, A_W (\tau_W) + C_t^\gamma \frac{4}{3}  A_t (\tau_t)\; 
[\kappa_t+\kappa_{\gamma \gamma} ] }{ A_W (\tau_W) +  C_t^\gamma \frac{4}{3} A_t (\tau_t)} \right|\,.
\eeq
These formulas show clearly the correlation between the parameters. This correlation is absent in our proposed parameterisation.

\section{Data analysis}

The data analysed by the two collaborations (ATLAS and CMS) is presented in terms of measured cross sections  in the relevant decay channels and for various selection rules: in the following, we will make use of as much information as it is available.
For instance, for the $H \to \gamma \gamma$ channel, the measured signal is available for several selection cuts, while for $H \to ZZ \to 2\times (l^+ l^-)$ and $H \to W W$, only the total measured cross section is available.
In the case of the $WW$ channel, the searches are actually performed in various selection channels, however the detailed efficiency of each signal region for the various production mechanisms is not publicly available.
As in our parameterisation the contribution of the production channels to the total cross section is modified with respect to the SM, a meaningful fit is not possible without detailed information on the efficiencies. In a first step we will focus on the channels where all needed information is provided, $\gamma\gamma,\ ZZ,\ \bar bb$, and we will afterwards present a method to include also the $WW$ and $\bar\tau\tau$ channels.

In general, for each selection channel $i$ of the experiments, the data being fitted is represented by the best fit values of the signal strength $\hat{\mu_i}$ (defined as the observed number of events divided by the expected number of events for a SM Higgs boson) as well as its uncertainty $\sigma_i$. As described in \cite{AzatovContinoGalloway}, to compare these values to theoretical 
expectations, the signal strength in one channel must be compared to the one calculated in each model, given for a specific 
channel by:
\begin{equation}
\mu_i = \frac{n_s^i}{(n_s^i)^{SM}}=\frac{\sum_p \sigma_p\ \epsilon_p^i}{\sum_p \sigma_p^{SM} \epsilon_p^i} \times \frac{BR_i}{BR_i^{SM}}\,,
\label{eq:mui}
\end{equation}
where $n_s^i$ is the predicted number of signal events in channel $i$ in the studied model, and $(n_s^i)^{SM}$ that same number in 
the SM. For each production mode $p$ (theoretical calculation at NLO) the efficiency of selection of a channel $i$ (experimental observation) 
is given by $\epsilon_p^i$, considered to stay the same with new physics. 
This assumption is in agreement with our parameterisation that only includes corrections that do not change the kinematics of the events.
At last, $BR_i$ and $BR_i^{SM}$ are the branching ratios of the 
Higgs boson into the decay channel corresponding to the selection channel $i$, for both SM and studied model.

The following data have been used for each channels :
\begin{itemize}
 \item $H\to\gamma\gamma$ : in CMS, information was given in \cite{CMSgamgamnote}: in 
Table~2 of the reference, one can find the product $\hat{\sigma}_p^i=({\sigma_p^{SM} \epsilon_p^i})/({\sum_{p'} \sigma_{p'}^{SM} \epsilon_{p'}^i})$ for each selection 
and production channel, whereas the best fit values as well as the uncertainties 
can be found in another table on the corresponding TWiki~\cite{twiki}, for a Higgs mass of $m_H = 125$ GeV. Those results are shown in Table \ref{DataCMS}. The results from the ATLAS 
collaboration, on the other hand, can be found in note \cite{ATLASgamgamnote}, where the selection 
channel efficiencies are given in Table~6, and the best fit values had to be extracted from Figures 14a and 14b for 7 TeV and 8 TeV respectively. They are shown in Table \ref{DataATLAS}. 
  \item $H\to ZZ^*\to \ell^+\ell^-\ell^+\ell^-$ : in this case, no tables were given by either experiment and we had to assume identical cut efficiencies for all production channels. 
This seems to be a reasonable assumption for an inclusive channel like $ZZ$, however this may not be
the case if discriminants based on the kinematic properties of the leptons are used, as it is the
case for CMS, where different efficiencies for different production channels may arise.
The assumption of universal efficiency allowed us to replace the efficiency-scaled cross sections 
$\hat{\sigma}_p^i$ by standard cross sections $\sigma_p$, which were taken from the LHC Higgs Cross-section Working Group's 
Website \cite{HiggsPRBR}. Best fit values and uncertainties were extracted from Figure 19 in  
\cite{CMSZZnote}, and Figure 16a from the note \cite{ATLASZZnote}, and the results are presented in Table \ref{DataZZ}.
  \item $H\to \bar bb$ : The most sensitive part\footnote{This final state is indeed also looked for
in the $\bar ttH$ production, but the sensitivity is still low at the time being.} comes only
through VH production, thus one can set all other efficiencies to zero, and the efficiency of the VH
production cancels out in the ratio in Eq.~\ref{eq:mui}. The experimental results are
$\hat{\mu}=-0.4\pm1.1$ for ATLAS and $\hat{\mu}=1.3\pm0.7$ for CMS (\cite{CMSbbnote,ATLASbbnote}).
\end{itemize}

\begin{table}
\begin{center}
 \begin{tabular}{|l|c|c|c|c|c|c|c|c|c|c|}
  \hline
    Selection channel & $\ \hat{\sigma}_{gg}^i \ \rule[-7pt]{0pt}{20pt}$ & $\hat{\sigma}_{VBF}^i $ & $\ \hat{\sigma}_{VH}^i \ $ & $\ \hat{\mu}_i\ $ & $\ \sigma_i\ $ & $\ \hat{\sigma}_{gg}^i\ $ & $\hat{\sigma}_{VBF}^i $ & $\ \hat{\sigma}_{VH}^i \ $ &  $\ \hat{\mu}_i\ $ & $\ \sigma_i\ $ \\
  \cline{2-11}
	& \multicolumn{5}{c|}{7 TeV} & \multicolumn{5}{c|}{8 TeV} \\  \hline
Untagged 0 & 61 & 17 & 19 & 3.15 & 1.82 & 68 & 12 & 16 & 1.46 & 1.24 \\  
\hline
Untagged 1 & 88 & 6 & 6 & 0.66 & 0.95 & 88 & 6 & 6 & 1.51 & 1.03  \\
  \hline
Untagged 2 & 91 & 4 & 4 & 0.73 & 1.15 & 92 & 4 & 3 & 0.95 & 1.15\\
  \hline
Untagged 3 & 91 & 4 & 4 & 1.53 & 1.61 & 92 & 4 & 3 & 3.78 & 1.77 \\
  \hline
Dijet Tag & 27 & 73 & 1 & 4.21 & 2.04 & - & - & - & - & - \\
  \hline
Dijet tight & - & - & - & - & - & 23 & 77 & 0 & 1.32 & 1.57 \\
  \hline
Dijet loose & - & - & - & - & - & 53 & 45 & 2 & -0.61 & 2.03 \\
  \hline
 \end{tabular}
\caption{CMS results in the $H \rightarrow \gamma \gamma$ channel \cite{CMSgamgamnote,twiki}.}
\label{DataCMS}
\end{center}
\end{table}

\begin{table}
\begin{center}
 \begin{tabular}{|p{3 cm}|c|c|c|c|c|c|c|c|c|c|c|c|}
  \hline
    Selection channel & $\hat{\sigma}_{gg}^i \rule[-7pt]{0pt}{20pt}$ & $\hat{\sigma}_{VBF}^i $ & $\hat{\sigma}_{WH}^i $ & $\hat{\sigma}_{ZH}^i $ & $\hat{\mu}_i$ & $\sigma_i$ & $\hat{\sigma}_{gg}^i$ & $\hat{\sigma}_{VBF}^i $ & $\hat{\sigma}_{WH}^i $ & $\hat{\sigma}_{ZH}^i $ &  $\hat{\mu}_i$ & $\sigma_i$ \\
  \cline{2-13}
	& \multicolumn{6}{c|}{7 TeV} & \multicolumn{6}{c|}{8 TeV} \\
  \hline
Unconverted central low $p_{Tt}$ & 92.9  & 4.0 & 1.8 & 1 & 0.5 & 1.4 & 92.9  & 4.2 & 1.7 & 1.0 & 1.0 & 1.2 \\
  \hline
Unconverted central high $p_{Tt}$ & 66.5  & 15.7 & 9.9 & 5.7 & 0.2 & 1.9 & 72.5  & 14.1 & 6.9 & 4.2 & 0.8 & 1.7 \\
  \hline
Unconverted rest low $p_{Tt}$ & 92.8  & 3.9 & 2.0 & 1.1 & 2.4 & 1.6 & 92.5  & 4.1 & 2 & 1.1 & 0.9 & 1.4 \\
  \hline
Unconverted rest high $p_{Tt}$ & 65.4  & 16.1 & 10.8 & 6.1 & 10.3 & 3.8 & 72.1  & 13.8 & 7.8 & 4.6 & 1.8 & 1.8 \\
  \hline
Converted central low $p_{Tt}$ & 92.8  & 4.0 & 1.9 & 1.0 & 6.2 & 2.6 & 92.8  & 4.3 & 1.7 & 1.0 & 3.4 & 2.0 \\
  \hline
Converted central high $p_{Tt}$ & 66.6  & 15.3 & 10 & 5.7 & -4.4 & 1.6 & 72.7  & 13.7 & 7.1 & 4.1 & 3.5 & 2.7 \\
  \hline
Converted rest low $p_{Tt}$ & 92.8  & 3.8 & 2.0 & 1.1 & 2.7 & 2.2 & 92.5  & 4.2 & 2 & 1.1 & 0.4 & 1.8 \\
  \hline
Converted rest high $p_{Tt}$ & 65.3  & 16.0 & 11.0 & 5.9 & -1.7 & 3 & 70.8  & 14.4 & 8.3 & 4.7 & 0.3 & 2.1 \\
  \hline
Converted transition & 89.4  & 5.2 & 3.3 & 1.7 & 0.3 & 3.7 & 88.8  & 6.0 & 3.1 & 1.8 & 5.5 & 3.3 \\
  \hline
2 jets & 22.5  & 76.7 & 0.4 & 0.2 & 2.7 & 1.9 & 30.4  & 68.4 & 0.4 & 0.2 & 2.6 & 1.8 \\
\hline
 \end{tabular}
\caption{ATLAS results in the $H \rightarrow \gamma \gamma$ channel \cite{ATLASgamgamnote}.}
\label{DataATLAS}
\end{center}
\end{table}

\begin{table}
\begin{center}
 \begin{tabular}{|l|c|c|c|c|c|c|c|}
  \hline
    Experiment & $\ \ \sigma_{gg} [pb]\ \ \rule[-7pt]{0pt}{20pt}$ & $\ \ \sigma_{VBF} [pb]\ \ $ & $\ \ \sigma_{WH} [pb]\ \  $ & $\ \ \sigma_{ZH} [pb]\ \ $ & $\sigma_{tth} [pb]$ & $\ \ \ \ \hat{\mu}\ \ \ \ $ & $\ \ \ \ \sigma\ \ \ \ $ \\
\hline
  CMS  $ZZ$ & \multirow{2}{*}{15.32} & \multirow{2}{*}{1.222} & \multirow{2}{*}{0.5729} & \multirow{2}{*}{0.3158} & \multirow{2}{*}{0.08634} & 0.8 & 0.35 \\
   \cline{1-1}
  \cline{7-8}
 ATLAS  $ZZ$ &  &  &	&  &     & 1.3 & 0.6 \\
  \hline
 \end{tabular}
\caption{CMS and ATLAS results in the $H \rightarrow ZZ \rightarrow \ell^+\ell^-\ell^+\ell^-$
channel \cite{CMSZZnote,ATLASZZnote}.}
\label{DataZZ}
\end{center}
\end{table}

When combining with our procedure the $\gamma \gamma$ ATLAS results, we met a
problem with selection channel \textit{Converted central high $p_{Tt}$}. Whereas for all 
the other channels the combination of 7 TeV and 8 TeV data gave results in agreement with the ones reported in Figure 14c in \cite{ATLASgamgamnote}, it was 
not the case for this channel. 
We therefore decided to ignore it in 7 TeV as well as 8 TeV data: the following ATLAS
plots are combination of all remaining channels.

To fit our parameterisation, we used a $\chi^2$ method where each computed signal strength is compared to its corresponding 
data best fit value through
\begin{equation}
\chi^2 = \sum_i \frac{\left(\mu_i-\hat{\mu}_i\right)^2}{\sigma_i^2}\,.
\label{eq:chi2}
\end{equation}
The values of $\chi^2$ computed this way are then compared to the exclusion thresholds at 68 and 95\% CL for a $m$ degrees of 
freedom $\chi^2$ distribution, where $m$ is the number of channels $i$ minus the number of independent fitted parameters (i.e. the number of independent parameters which value are fitted to the data).

\subsection{Improved $\chi^2$ method}
\label{sec:imp_chi}
As mentioned previously, the $\chi^2$ obtained in Eq.~\ref{eq:chi2} meets two serious obstacles:
first, one needs the efficiency (or equivalently the signal sample composition) per production mode
together with the best fit $\hat{\mu}_i$ in each sub-channel and, second, this procedure neglects the
correlations between uncertainties of the different sub-channels. However, it is possible to go
further by recasting the results of the couplings analyses made by the two collaborations
\cite{cms_comb,atlas_comb}. Indeed, instead of providing the best fits as one dimensional
distributions ($\hat{\mu}_i\pm \sigma_i$), we now have access to two dimensional distributions
($(\hat{\mu}_{i,ggH/\bar ttH},\hat{\mu}_{i,VBF/VH})$, with the one sigma contour). Under some
acceptable assumptions, this information can be used as the $\chi^2_i$ function of the channel $i$,
without explicit reference to sub-channels and efficiencies. Those assumptions are the following :
\begin{itemize}
 \item VBF and VH production are rescaled in the same way. This is achieved by imposing $\kappa_Z=\kappa_W$, which is the case in any model respecting the custodial symmetry.
 \item For each channel $i$, there cannot be significant contributions from both the gluon fusion and
the $t$ quark associated production. At the time being, this is the case, since $\bar ttH$ is
significant only in the $H\to\bar bb$ channel which is otherwise observable only through VH.
\end{itemize}
As for the previous method, we must also assume that the Gaussian approximation is well motivated. We can then write the approximated likelihood of a given channel $i$ as
\begin{equation}
 \chi^2_i=\binom{\mu_{i,ggH/\bar ttH}-\hat{\mu}_{i,ggH/\bar
ttH}}{\mu_{i,VBF/VH}-\hat{\mu}_{i,VBF/VH}}^TV_i^{-1}\binom{\mu_{i,ggH/\bar
ttH}-\hat{\mu}_{i,ggH/\bar ttH}}{\mu_{i,VBF/VH}-\hat{\mu}_{i,VBF/VH}} \,,
 \label{eq:chi2_imp}
\end{equation}
which is nothing but the 2D version for Eq.~\ref{eq:chi2}, where $V_i$ and $\hat{\mu}_{i,X}$ are
obtained by fitting the one sigma contour to $\chi^2_i(1\sigma)=2.3$ (this value correspond to the
68\% CL of a two-dimensional $\chi^2$ distribution), see figures in Table \ref{tab:ellipses}. The
main difference with the 1D version is that
we do not get directly the $\chi_i^2$ function, which \textit{a priori} does not vanish anywhere,
but its deviation to the best fit $\Delta \chi_i^2$ which vanishes at the best fit point. However we
have checked that the resulting statistical test yields a conservative result as compared to the
true $\chi^2$ test, as long as each best value $\chi^2_i(\hat{\kappa})$ had a large \textit{p}-value
(\textit{i.e.} not negligible as compared to one).

\begin{table}
\begin{center}
 \begin{tabular}{|l|c|c|}
  \hline
    Canal & $ \left(\hat{\mu}_{ggH/\bar ttH},\ \hat{\mu}_{i,VBF/VH}\right)
$ & $V$  \\
  \hline
    $H \to \gamma\gamma$ & $\left(0.95,\ 3.77\right) $&
$\left(\begin{array}{cc} 0.95 & -1.35 \\ -1.35 & 6.87 \end{array} \right)$ \\
  \hline
    $H \to WW$ & $\left(0.77,\ 0.39 \right) $&$
\left(\begin{array}{cc} 0.19 & 0.15 \\ 0.15 & 1.79 \end{array} \right)$ \\
  \hline
    $H \to \tau\tau$& $\left(0.93,\ 0.89\right) $&
$\left(\begin{array}{cc} 2.02 & -0.92 \\ -0.92 & 2.14 \end{array} \right)$ \\
  \hline
 \end{tabular}
\caption{CMS results in the $H \rightarrow WW$, $H \to \tau\tau$ and $H \rightarrow \gamma\gamma$
channels. Results inferred from \cite{cms_comb}, using equation \ref{eq:chi2_imp}.}
\label{tab:ellipses}
\end{center}
\end{table}

We have mainly used this improved $\chi^2$ for the CMS results, since all\footnote{Of course there
is no need for such a treatment in the $ZZ$ channel, since it does not distinguish productions
modes.} channels could be treated this way, whereas the update of the $WW$ channel in ATLAS did not
include this information. Given that the uncertainties are so far statistically dominated, they are
mostly uncorrelated hence we do not expect significant differences with the previous method, which
we have checked by comparing the two methods on the same $H\to\gamma\gamma$ analysis from CMS.
However this method is crucial for channels where efficiencies are not available (for instance the
$WW$) and it may also be promising when the correlations become important in uncertainties.\\

In the following sections, we will use our reconstructed likelihood to derive
confidence regions in our
parameter space. This requires the choice of a statistical test, and we have compared two
different tests. The first is the profiled likelihood ratio, \textit{i.e.} the quantity
$\Delta\chi^2(\kappa)=\chi^2(\kappa)-\chi^2(\hat{\kappa})$ and the second is the full $\chi^2$
test. They differ in the sense that the $\Delta\chi^2$ test is the assessment of a given hypothesis
($\kappa$) as compared to another hypothesis ($\hat{\kappa}$), whereas the $\chi^2$ test assesses
the $\kappa$ hypothesis without reference to other hypotheses. Depending on the data, they will
have a different power: in general the $\Delta\chi^2$ will be stronger as long as the best fit
$\hat{\kappa}$ is sufficiently likely; if this is not the case, then the $\chi^2$ test should be used.
This can be
understood as the fact that the $\Delta\chi^2$ does not test whether a given choice of the parameters $\kappa$ suitably describes the data, however it tests how a given
point
in the parameter space compares to the best fit point $\hat{\kappa}$. In the following sub-sections we present some sample fits, where we have
chosen the most appropriate test case by case.

\subsection{Simple two parameter fit}

To start with, one can restrict the fit to the two parameters describing the loop couplings only.
This can be justified in two cases:
\begin{itemize}
\item[-] new physics only enters via loops, while corrections to tree level couplings are small;
\item[-] the most sensitive measurements only involve loop induced couplings, while the effects on tree level processes are subleading.
\end{itemize}
In the second case, one can absorb the contribution of $\kappa_t$ and $\kappa_W$ into the loop parameters:
\beq
\kappa'_{gg} &=& \kappa_{gg} + \kappa_t-1\,, \\
\kappa'_{\gamma \gamma} &=& \kappa_{\gamma \gamma} + \kappa_t-1 + \frac{3 A_W}{4 A_t} (\kappa_W-1)\,.
\eeq
Note for example that the $t\bar{t}h$ coupling is not measurable at the moment, thus $\kappa_t$ can always be absorbed in 
$\kappa_{gg}$ and $\kappa_{\gamma \gamma}$.
Furthermore, a small contribution to $\kappa_W-1$ can generate sizeable effects on $\kappa_{\gamma \gamma}$ due to the enhancement factor $\frac{3 A_W}{4 A_t} \sim -4.6$. 
In practice, present experimental results allow to perform a meaningful fit only with a very restricted set of parameters.
This relation is also in agreement with the expectation that modifications to the tree level couplings, $\kappa_W$ and $\kappa_t$, are generated by New Physics effects at tree level.

In the two parameter case \cite{Cacciapaglia:2009ky}, the signal strength can be calculated with all given data, as the cross 
sections for each production channel is proportional to the one in the SM. We can therefore write for the
$H \rightarrow \gamma \gamma$ channels:
\begin{eqnarray}
\mu_{i,\ \gamma\gamma} & = & \frac{(1+\kappa_{gg})^2 \sigma_{gg}^{SM} \epsilon_{gg}^i +
\sigma_{VBF}^{SM}\epsilon_{VBF}^i + \sigma_{ZH}^{SM}\epsilon_{ZH}^i + \sigma_{WH}^{SM}\epsilon_{WH}^i}
{\sigma_{gg}^{SM}\epsilon_{gg}^i + \sigma_{VBF}^{SM}\epsilon_{VBF}^i + \sigma_{ZH}^{SM}\epsilon_{ZH}^i + \sigma_{WH}^{SM}\epsilon_{WH}^i}\times \frac{BR_{\gamma \gamma}}{BR_{\gamma \gamma}^{SM}} = \nonumber\\
  & = & \frac{1+ \left[ (1+\kappa_{gg})^2 -1 \right]\hat{\sigma}_{gg}^i}{1+ \left[ (1+\kappa_{gg})^2 -1 \right]BR^{SM}_{gg}} \left(1+\frac{\kappa_{\gamma\gamma}}{\frac{9}{16}A_W(\tau_W)+1}\right)^2\,.
\end{eqnarray}
The factor containing the SM branching $BR_{gg}^{SM}$ takes into account the change in the Higgs total width due to the modification to the gluon couplings (while we neglect the effect of the photon channel in the total width). 
For other channels, we also finds similar equations with different efficiencies and branching
ratios, as for instance the $H \rightarrow ZZ \rightarrow \ell^+\ell^-\ell^+\ell^-$ channel:
\begin{equation}
\mu_{ZZ} = \frac{(1+\kappa_{gg})^2 +
\frac{\sum_{oth}{\sigma_{oth}}}{\sigma_{gg}}}{1 +
\frac{\sum_{oth}{\sigma_{oth}}}{\sigma_{gg}}} \frac{1}{1+ \left[ (1+\kappa_{gg})^2 -1
\right]BR^{SM}_{gg}}\,.
\end{equation}

\begin{figure}[tb]
\begin{center}
\includegraphics[width=0.45\textwidth]{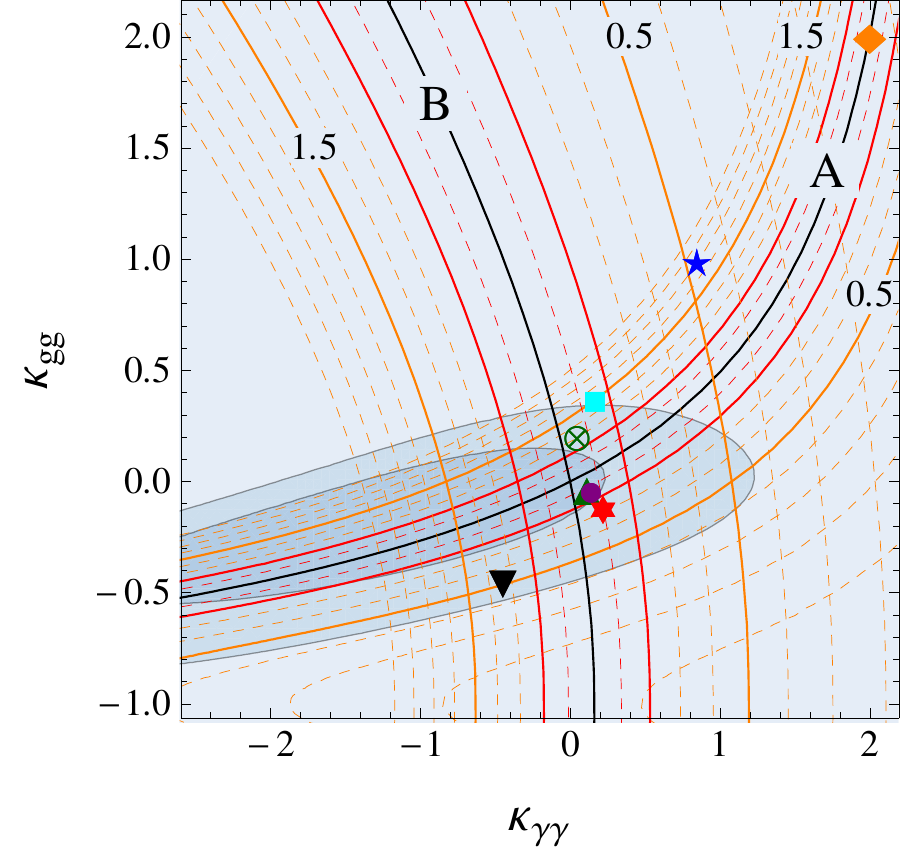}~~~~~
\includegraphics[width=0.45\textwidth]{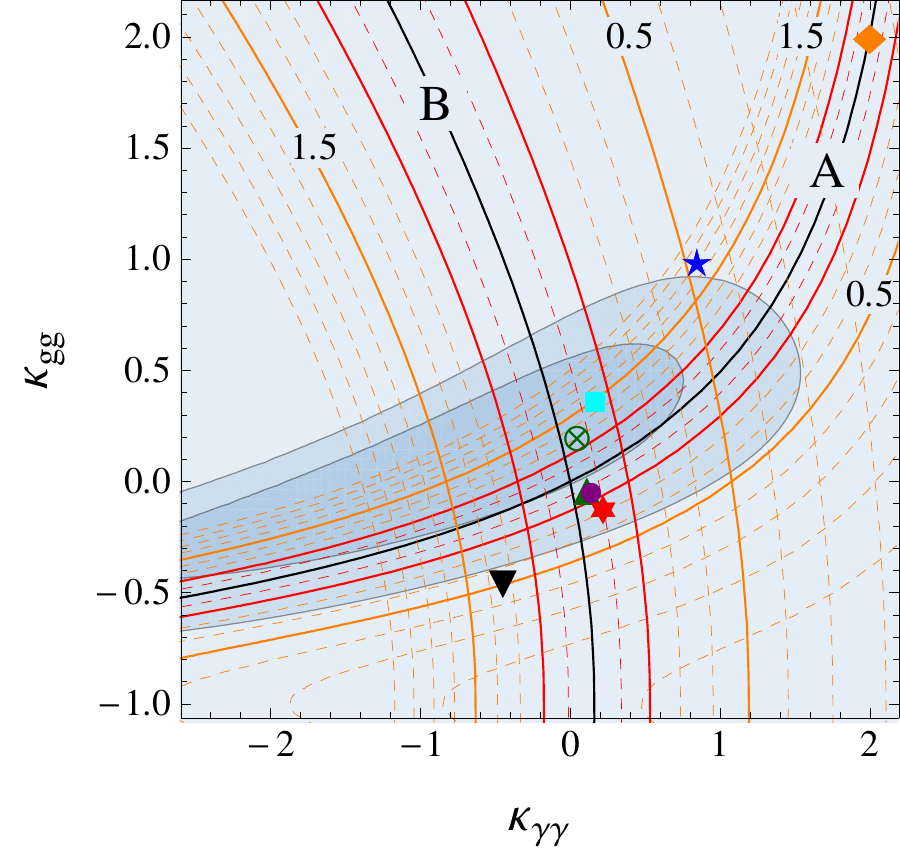}
\end{center}
\caption{\footnotesize  $\kappa'_{\gamma \gamma}$ and $\kappa'_{g g}$ at the LHC for a  Higgs boson with $m_H = 125$ GeV.
The two solid lines correspond to the SM values of the inclusive $\gamma \gamma$ channel ({\bf A}), and the vector boson fusion 
production channel ({\bf B}). On the left panel, the fit using ATLAS data. On the right, the fit
using CMS data. Both fits use $\gamma\gamma$, $ZZ$ and $\bar bb$ channels. Darker (lighter) blue
are 
the 1, 2 $\sigma$ limits. 
}
\label{fig:fitlight}
\end{figure}

Using the published CMS and ATLAS results, the result of the two parameter fit using $\kappa_{gg}$ and $\kappa_{\gamma \gamma}$ is given 
in Figure \ref{fig:fitlight} using data from $\gamma\gamma$, $ZZ$ and $\bar bb$ channels. One can
see that the allowed region runs along the line $A$, which corresponds to SM cross section for the process $gg \to h \to \gamma \gamma$, even though largish values of $\kappa_{gg}$ are
excluded by the $ZZ$ channel. As can be seen on the different contours, the current data already
imposes a constraint on possible departures from the Standard Model point.

For reference, sample points for the following models are indicated:
\begin{itemize}
\item[-] [\textcolor{orange}{$\blacklozenge$}] fourth generation, where the result is independent on the masses and Yukawa couplings. The point is given here for illustrative purposes; more complete fits on this case exist, 
see for example \cite{Djouadi:2012ae,Eberhardt:2012gv};
\item[-] [\textcolor{red}{$\ast$}] Littlest Higgs~\cite{LH}, where the result scales with the symmetry breaking scale $f$, set here to $f=500$ GeV for a model with $T$-parity (there is also a mild dependence on the triplet VEV $x$, that we set $x=0$);
\item[-] [\textcolor{vertfonce}{$\blacktriangle$}] Simplest Little Higgs~\cite{SLH} (see Section~\ref{sec:SLH}), where the result scales with the $W'$ mass, also set here to $m_{W'}=500$ GeV for a model with $T$-parity;
\item[-] [\textcolor{cyan}{$\blacksquare$}] colour octet model~\cite{octet}, where the result is inversely proportional to the mass $m_S = 750$ GeV in the example (and also depends on two couplings set here to $\lambda_1 = 4$, $\lambda_2 = 1$);
\item[-] [\textcolor{vertfonce}{$\otimes$}] 5D Universal Extra Dimension model~\cite{UED}, where only the top and $W$ resonances contribute and the result scales with the size of the extra dimension (here we set $m_{KK} = 500$ GeV close to the experimental bound from electroweak precision tests);
\item[-] [\textcolor{blue}{$\bigstar$}] 6D UED model on the Real Projective Plane~\cite{RPP}, with $m_{KK}= 600$ GeV is set to the LHC bound~\cite{RPPb};
\item[-] [\textcolor{purple}{$\bullet$}] the Minimal Composite Higgs~\cite{GHUwarped} (Gauge Higgs unification in warped space) with the IR brane at $1/R' = 1$ TeV, where only $W$ and top towers contribute significantly and the point only depends on the overall scale of the KK masses, as the other parameters are fixed by the $W$ and top masses;
\item[-] [\textcolor{gray}{$\blacktriangledown$}] a flat ($W'$ at 2 TeV) and [\textcolor{pink}{$\spadesuit$}] warped ($1/R'$ at 1 TeV) version of brane Higgs models, in both cases the hierarchy in the fermionic spectrum is explained by the localisation, and all light fermion towers contribute; notwithstanding the many parameters in the fermion sector, the result only depends on the overall scale of the KK masses.
\end{itemize}
The numerical values of the parameters, whether including only $\kappa_t$ or both $\kappa_t$ and
$\kappa_W$ in the loop parameters, are given in Table~\ref{tab:models} for all the models listed
above.
The values are computed using the results in~\cite{Cacciapaglia:2009ky}, while the 6D UED model has been computed in~\cite{oda}.
Note that the fit we present here is based on two independent parameters, while in many of the models we show all the parameters depend on a single model parameter, typically the mass scale of the new physics.
Therefore, one should take any effective parameter fit with the caveat that the fit must be redone for a specific model taking into account the actual number of independent parameters in the BSM model.
Note also that in all cases, except the fourth generation, the result scales with the mass of the
new particles, therefore a point that falls in the exclusion region implies that a stronger bound on
the mass of the new states is imposed by the Higgs measurements rather that the exclusion of the
model.
We should therefore think of the model as covering a line of points connecting the benchmark point in the Figures with the origin (SM point).
\begin{table}
\begin{center}
 \begin{tabular}{|l|l|c|c|c|c|}
  \hline
    Model & parameter(s) & $\kappa_W-1$   & $\kappa'_{gg} (\kappa_t)$ & $\kappa'_{\gamma \gamma} (\kappa_t)$ & $\kappa'_{\gamma \gamma} (\kappa_t, \kappa_W)$ \\
\hline
  4$^{\rm th}$ generation & - & 0 & 2 & 2 & 2 \\
\hline
  Simplest Little Higgs & $m_{W'} = 500$ GeV & -0.009 & -0.034 & 0.067 & 0.11 \\
  \hline
  Littlest Higgs & $f = 700$ GeV & -0.05 & -0.11 & -0.014 & 0.23 \\
                       & $m_{W'} = 500$ GeV, $x=0$ &   &   &   &  \\
  \hline
  colour octet & $m_S = 750$ GeV & 0 & 0.37 & 0.17 & 0.17 \\
                      & $\lambda_1 = 4$, $\lambda_2 = 1$ &  &  &  &  \\
  \hline
    5D UED & $m_{KK} = 500$ GeV & 0 & 0.20 & 0.034 & 0.034 \\
  \hline
    6D UED (RP$^2$) & $m_{KK} = 600$ GeV & 0 & 1.00 & 0.84 & 0.84 \\
                                &    ($R_5 = 1.5\; R_4$)     &    &     &    &   \\
  \hline
    composite Higgs & $1/R' = 1$ TeV & -0.04 & -0.04 & -0.03 & 0.14 \\
  \hline
    flat brane Higgs & $m_{W'} = 2$ TeV & -0.005 & -0.45 & -0.47  & -0.45 \\
  \hline
    warped brane Higgs & $1/R' = 1$ TeV & -0.11 & -0.65 & -1.08 & -0.57 \\
  \hline
 \end{tabular}
\caption{Higgs coupling parameters for various benchmark models: in parenthesis we indicate if
$\kappa_t$ and/or $\kappa_W$ are included in the definition of the loop parameters. In the second
column, the mass parameter 	the corrections are inversely proportional to (eventual other
parameters are indicated in parenthesis). Here we only consider corrections to $\kappa_W$ generated
at tree level.}
\label{tab:models}
\end{center}
\end{table}

\begin{figure}[!t]
\begin{center}
\includegraphics[width=0.45\textwidth]{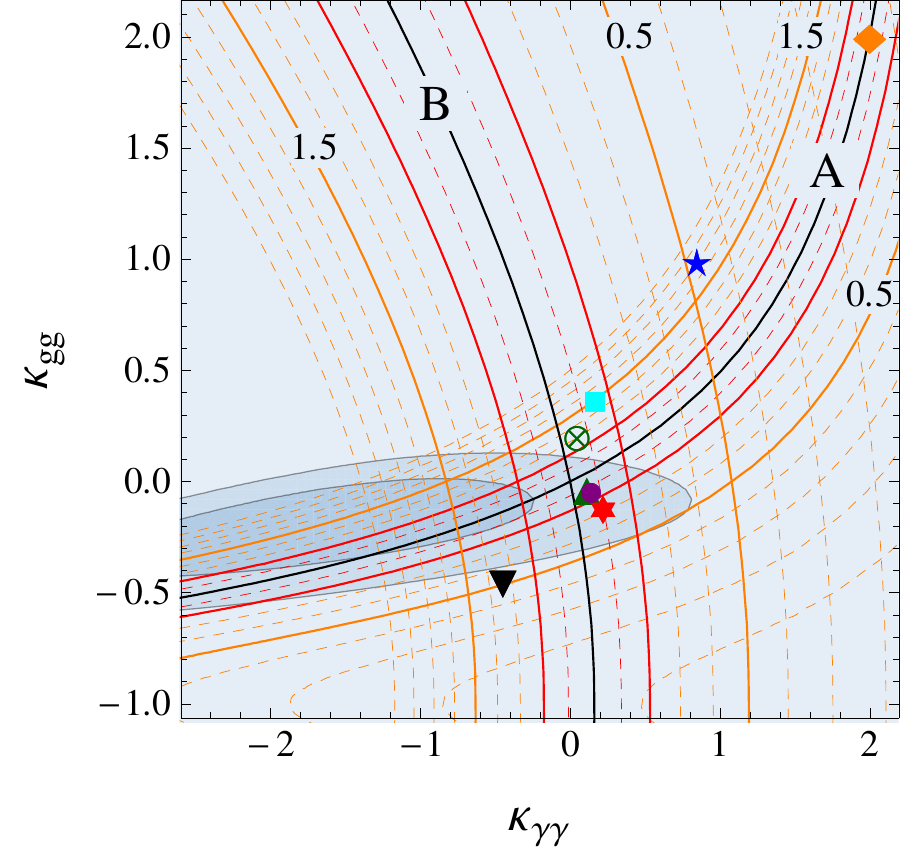}
\end{center}
\caption{\footnotesize $\kappa'_{\gamma \gamma}$ and $\kappa'_{g g}$ at the LHC for a  Higgs boson
with $m_H = 125$ GeV. This plot use CMS data from all channels.}
\label{fig:CMS_2p_alldata}
\end{figure}

The fit in Figure \ref{fig:fitlight} has been done using the $\chi^2$ in Eq. \ref{eq:chi2}.
In order to include the $WW$ and $\bar\tau\tau$, we now use the improved $\chi^2$.
Thus we will focus on CMS
data, and use $ZZ$ and $\bar bb$ with likelihoods from Eq.~\ref{eq:chi2} (since we assume negligible
change in $\bar ttH$ production, we can treat both channels as one dimensional distributions over
the production modes) and $WW$, $\gamma\gamma$ and $\bar\tau\tau$ with likelihoods from
Eq.~\ref{eq:chi2_imp}. We show the resulting contours in Figure \ref{fig:CMS_2p_alldata}. By reducing
to a two parameter fit on $\kappa_{gg}',\kappa_{\gamma\gamma}'$ we make the assumption that $k_Z$ is
close to one, and that the changes in $k_t,k_W$ are noticeable only in the loops, not in direct
production or decays. All specific models fall into this category but the warped brane Higgs, which
deviate significantly in $k_W$. To test this model, we will need to go beyond the 2-parameter fit.

\subsection{Three parameter fit}

As a sample and more general application of our parameterisation, we have also performed a three parameter fit, by adding to the two previous variables an explicit dependence on the gauge boson couplings.
For simplicity, we assume that a custodial symmetry present in the BSM models imposes $\kappa_Z = \kappa_W = \kappa_V$, so that we can treat deviations from the SM couplings of both $W$ and $Z$ with a single parameter.
This is usually the case in any reasonable model of New Physics.

In the 3 parameter fit, the expression of the signal strengths in the $\gamma \gamma$ channel can be written as
\begin{eqnarray}
\mu_{i,\ \gamma\gamma} & = & \frac{(1+\kappa_{gg})^2 \sigma_{gg}^{SM} \epsilon_{gg}^i + \kappa_V^2 \left(
\sigma_{VBF}^{SM}\epsilon_{VBF}^i + \sigma_{ZH}^{SM}\epsilon_{ZH}^i + \sigma_{WH}^{SM}\epsilon_{WH}^i \right)}
{\sigma_{gg}^{SM}\epsilon_{gg}^i + \sigma_{VBF}^{SM}\epsilon_{VBF}^i + \sigma_{ZH}^{SM}\epsilon_{ZH}^i + \sigma_{WH}^{SM}\epsilon_{WH}^i}\times \frac{BR_{\gamma \gamma}}{BR_{\gamma \gamma}^{SM}} =\\
  & = & \frac{(1+\kappa_{gg})^2 \hat{\sigma}_{gg}^i + \kappa_V^2 (1-\hat{\sigma}_{gg}^i)}{1+ (\kappa_V^2-1) (BR^{SM}_{WW+ZZ}) +  \left[ (1+\kappa_{gg})^2 -1 \right]BR^{SM}_{gg}} \left(1+\frac{\kappa_{\gamma\gamma}}{\frac{9}{16}A_W(\tau_W)+1}\right)^2\,. \nonumber
\end{eqnarray}
For the $ZZ$ channel:
\begin{equation}
\mu_{ZZ} = \frac{(1+\kappa_{gg})^2 \sigma_{gg} + \kappa_V^2 \sum_{oth}{\sigma_{oth}}}{\sigma_{gg} +
\sum_{oth}{\sigma_{oth}}} \frac{\kappa_V^2}{1+ (\kappa_V^2-1) (BR^{SM}_{WW+ZZ}) +  \left[ (1+\kappa_{gg})^2 -1 \right]BR^{SM}_{gg}}\,.
\end{equation}

\begin{figure}[tb]
\begin{center}
\includegraphics[width=0.45\textwidth]{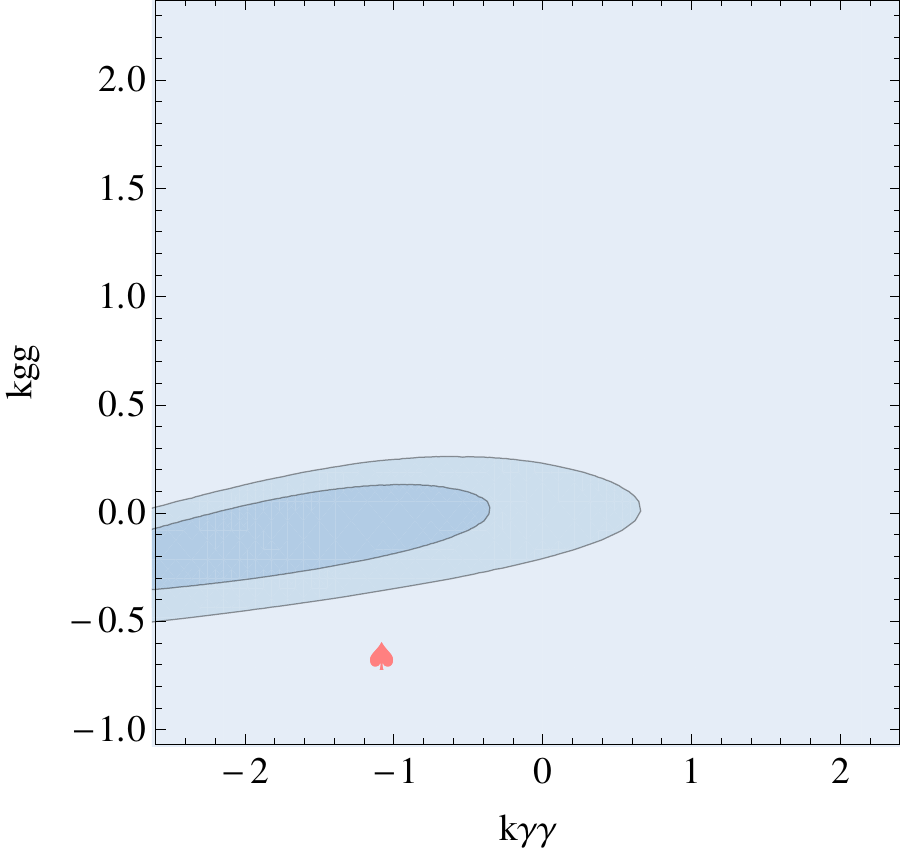}
\end{center}
  \caption{\footnotesize  Three parameter fit at the LHC for a Higgs boson with $m_H = 125$ GeV using
  all channels from CMS.
  Here we present a slice of the allowed region for $\kappa_V = 0.89$. Darker (lighter) blue are 
  the 1, 2 $\sigma$ limits. }
\label{fig:fitlight3}
\end{figure}

In Figure~\ref{fig:fitlight3} we present the results of the 3 parameter fit, by slicing the allowed
region for $\kappa_V = 0.89$.
This choice is motivated by the fact that this value correspond to the brane Higgs model in
Table~\ref{tab:models}. Thus we are taking the precise slice where the model lies, but one should
not forget that we are dealing with a three parameter space. In particular the interpretation is
different from a marginalisation over $k_V$. We see that the allowed region is very close to the one
obtained in the 2 parameter fit in Figure~\ref{fig:CMS_2p_alldata}.

\section{Model interpretations}

In general, the contribution of the new loops to the parameters can be written as~\cite{Cacciapaglia:2009ky}:
\beq
\kappa_{\gamma \gamma} &=& \sum_{NP}\;  \frac{C_{NP}^\gamma}{C_t^\gamma}\,  \frac{3}{4} N_{c,NP} Q_{NP}^2\;  
g_{hNP}\; \epsilon_{NP}\,, \label{kappagamma}\\
\kappa_{g g} &=& \sum_{NP}\;  \frac{C_{NP}^g}{C_t^g}\,  2 C (r_{NP})\;  g_{hNP}\; \epsilon_{NP}\,,
\label{kappaglu}
\eeq
where $C_{NP}^\gamma/C_t^\gamma$ and $C_{NP}^g/C_t^g$ are the NLO corrections normalised to the SM ones for 
the top in the $\gamma \gamma$-h and gluon-gluon-h triangle loop vertex,
$N_{c,NP}$ is the number of colour components of the new 
states running in the loop, $Q_{NP}$ is its electromagnetic charge, $C (r_{NP})$ is the Casimir of the colour representation of the 
new state ($C = 1/2$ for a fundamental), and $\epsilon_{NP}$ encodes the amplitude of the new physics loop.
For masses of the state in the loop larger than the Higgs mass threshold, the amplitudes quickly asymptotise to a constant number 
that only depends on the spin of the state: this is already true with a very good precision for the top amplitude.
Therefore, $\epsilon_{NP}$ can be simply approximated by a number dependent on the spin of the new states:
\beq
\epsilon_{NP} &=& 1\; \quad \mbox{for fermions}\,; \nonumber \\
\epsilon_{NP} &=& -21/4\; \quad \mbox{for vectors}\,;  \\
\epsilon_{NP} &=& 1/4\; \quad \mbox{for scalars}\,.\nonumber
\eeq
The coupling of the new states to the Higgs is expressed in terms of the Higgs VEV dependence of the new state mass:
\beq
g_{hNP} = \frac{v}{m_{NP}} \frac{\partial m_{NP} (v)}{\partial v}\,.
\eeq

Thus, in order to connect the new parameters $\kappa_{gg}$ and $\kappa_{\gamma \gamma}$ to new physics models, it is enough to know the quantum numbers (charge, colour representation and spin) and the Higgs VEV dependent mass of the new particles.
Eq.s~\ref{kappagamma} and \ref{kappaglu} also allow to easily draw the correlation between the two parameters in specific models.
For instance, in models with a single new particle:
\beq
\frac{\kappa_{\gamma \gamma}}{\kappa_{gg}} = \frac{3}{8} \frac{N_{c,NP} Q_{NP}^2}{C (r_{NP})}\,,
\eeq
where we have assumed that the QCD corrections to the amplitudes are the same for the top and for new physics.
For a top partner (i.e. a particle with the same quantum numbers as the top, except the spin):
\beq\kappa_{\gamma \gamma} = \kappa_{gg}\,.
\eeq

In most of the models listed in Table~\ref{tab:models}, the result for the relevant parameters
depends on a single mass scale (with just mild dependence on other parameters of the model).
Therefore, the prediction cannot in principle be directly compared with any 2 or 3 (or more)
parameter fit. For instance, when using the $\Delta\chi^2$ statistical test, one should compared it
to a single parameter $\chi^2$ distribution. However, the results obtained with an effective
parameterisation, like the ones advocated here, can be used to give an indication if the data
favours or disfavours the BSM model under consideration.

\subsection{Higher-order corrections}

It is a well known fact that Higgs physics (in the Standard Model as well as in many of its
extensions) is largely affected by radiative corrections, in particular because of strong
interactions. However, the exact calculation of those corrections is \textit{a priori}
model-dependent, which makes it impossible to include corrections in an effective Lagrangian without adding new
parameters. 
In fact, higher order corrections can generate new operators or kinematic structures together with a simple re-scaling of the LO operators.

The parameterisation we propose, by splitting clearly the effect of tree level modification of tree level couplings and New Physics loop effects on the loop induced couplings, allows to easily add, at least partially, NLO corrections to the calculation. This is not a completely consistent procedure but the largest corrections are included.
For instance, we have already seen that NLO QCD corrections to the top loop contribution to both gg-h and $\gamma \gamma$-h vertices factorise and give a simple multiplicative factor.
We can expect that QCD corrections to the New Physics loops have the same structure and can be factorised: contributions of this type are already included in Eq.s~\ref{kappagamma} and \ref{kappaglu} via the factors $C_{NP}^\gamma$ and $C_{NP}^g$.
A precise and self-consistent calculation of such effects must be carried out in any given specific model.
In the BSM model predictions in Table~\ref{tab:models}, however, we assume that the QCD corrections are the same as in the SM top loop for simplicity.
This is true for the contribution of coloured fermions (which is the most common case) because the corrections are independent on the mass of the fermion.

As already mentioned, this parameterisation cannot include NLO corrections that generate Lorentz structures different from the LO SM ones.
For instance, electroweak 2-loop corrections to the gluon fusion cross sections, which are proportional to the coupling of the Higgs to $W$ and $Z$, cannot be included~\cite{Aglietti:2004nj}.
Other corrections are loops involving both the production and decay process: such effects, however, are expected to be small because we are considering processes with the production of a Higgs boson in an s-channel resonance.

Anyway, even in an effective theory approach based on effective operators the procedure to compute radiative corrections implies adding new parameters to deal with the new structures. Indeed full consistency of the procedure can be guaranteed but this requires inserting counterterms \cite{Passarino:2012cb}. From the point of view of a fit adding new parameters 
for describing a few per cent modification of the vertices is not necessarily an improvement. For example the two loop 
electroweak correction to $h \to \gamma \gamma$ is less than 2\%, while the effect on $gg \to h$ is of order 5\%.

\subsection{An example: the Simplest Little Higgs model} \label{sec:SLH}

The new parameterisation offers an easy interpretation in terms of new physics models.
The idea is that the contribution of new physics to the parameters will in general scale in a simple way with the mass of the new 
states: thus, any model will roughly cover a straight line originating at the origin (i.e. the SM point).
By simply measuring the length of the excluded line, one can extract the bound on the new physics mass scale.

As an example, we present here the case of the Simplest Little Higgs model, described in Section 3.3 of 
Ref.~\cite{Cacciapaglia:2009ky}. The model contains both a $W$ partner $W'$ and a top partner $T$.
Moreover, mixing between the two states generates a modification of the couplings of the Higgs to the SM $W$ and top.
Therefore, there will be a contribution to $\kappa_W$ and $\kappa_t$ from the modified tree level couplings, and to $\kappa_{gg}$ 
and $\kappa_{\gamma \gamma}$ from the $W'$ and $T$ loops.
From the $W$ sector, the contributions are:
\beq
\kappa_W = 1 - \frac{1}{3} \frac{m_W^2}{m_{W'}^2} \,, \qquad \kappa_{\gamma \gamma} (W') = \frac{63}{16} \frac{m_W^2}{m_{W'}^2}\,,
\qquad \kappa_{gg} (W') = 0\,.
\eeq
Here $m_{W'}$ is the mass of the $W'$ new particle.
From the top sector, we have:
\beq
\kappa_t = 1+\frac{m_t^2}{m_T^2} - \frac{4}{3} \frac{m_W^2}{m_{W'}^2}\,, \qquad \kappa_{gg} (T) = \kappa_{\gamma \gamma} 
(T) = - \frac{m_t^2}{m_T^2}\,.
\eeq
As $\kappa_t$ is not measurable, one can include its effects in the loop parameters (in this case, for illustration, we will leave $\kappa_W$ in the 
fit, so we are intending a 3 parameter fit):
\beq
\kappa'_{gg} (k_t, T) = - \frac{4}{3} \frac{m_W^2}{m_{W'}^2}\,, \qquad \kappa'_{\gamma \gamma} (k_t, T, W') = \frac{125}{48} 
\frac{m_W^2}{m_{W'}^2}\,, \qquad \kappa_W - 1 = - \frac{1}{3} \frac{m_W^2}{m_{W'}^2} \,.
\eeq
Thus:
\begin{itemize}
\item[-] the parameters of the fit only depend on $m_{W}/m_{W'}$ and scale like $1/m_{W'}^{2}$ (this scaling is lost in the $\kappa_g$ 
and $\kappa_\gamma$ parameterisation);
\item[-] the correlation between the three parameters (in this model) is explicit, 
$\kappa_{gg} = - \frac{64}{125} \kappa_{\gamma 
\gamma} = - 4 (\kappa_W-1)$.

\end{itemize}

This shows explicitly that specific model points (or regions) can be put in a global fit only for qualitative and illustrative purposes.
In order to exclude a particular model at a given confidence level the $\chi^2$ per degree of freedom should be calculate in the specific 
model, as in general this quantity is model dependent due to the fact that the supposedly independent fit parameters of a global 
analysis can be correlated in that specific model (in the example just given the three parameters reduce to only one independent 
parameter).

\subsection{Fermiophobic Higgs model}

In some cases, models are not represented by single points but still contain parameters that we can
directly cast in our parameterisation. For instance, we consider here a class of models in
which fermions do not couple to the Higgs boson and, therefore, all the Higgs phenomenology takes place via the couplings to vectors.
 As a consequence, the main decay channel $\bar bb$
disappears, as well as the main production mode, $gg \to h$ (occurring only through fermion loops).
These two combined effects leave the inclusive cross-sections for bosonic channels 
not too different from the SM expectations. The overall effect of this new physics is included in
the couplings of the Higgs boson to the W and Z bosons, through the coefficients 
$\kappa_W$ and $\kappa_Z$. All fermionic $\kappa_f$ are therefore set to zero,
together with $\kappa_{gg}$ and $\kappa_{\gamma \gamma}$.
We do not assume custodial symmetry in order
to be able to probe all possible models in the two dimensional $(\kappa_W,\kappa_Z)$ parameter space.\\

\begin{figure}[tb]
\begin{center}
\includegraphics[width=0.45\textwidth]{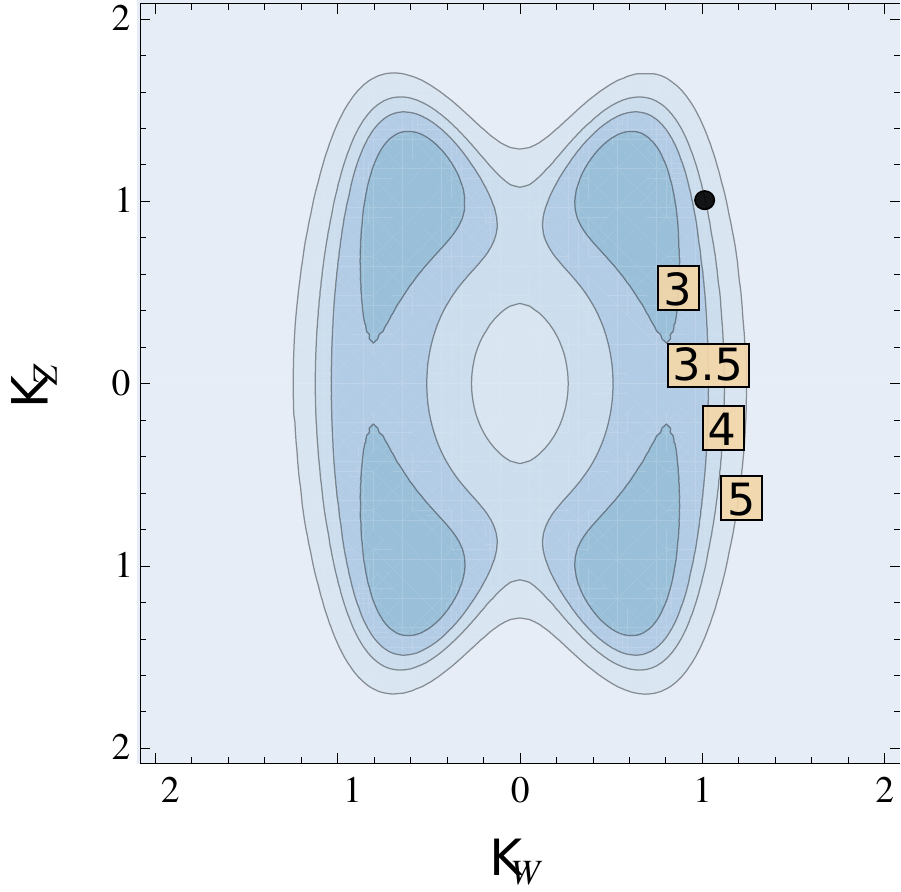}
\end{center}
\caption{\footnotesize Fermiophobic Higgs model fit at the LHC with $m_H = 125$ GeV using
all channels from CMS, in the $(\kappa_W,\ \kappa_Z)$ plane. Darker (lighter)
blue are 
the 3, 3.5, 4 and 5 $\sigma$ regions.  The black dot labels the fermiophobic SM, $\kappa_W= \kappa_Z=1$.}
\label{fig:fermiophobic}
\end{figure}

Apparently, this class of models precludes the use of the improved $\chi^2$ method described in
section \ref{sec:imp_chi}, since VBF and VH production are not rescaled in the same way. However it
is possible to include the $WW$ channel by noting that this channel is still quite
insensitive to VBF production mode (this is demonstrated in the CMS analysis \cite{cms_ww}). 
Concerning the $\bar\tau\tau$ channel, since this Higgs is fermiophobic, the signal is set to zero and therefore all production channels become irrelevant: in particular, we will have
$\mu_{VBF}=\mu_{VH}=0$ throughout the whole $(\kappa_W,\kappa_Z)$ plane, thus this satisfies the
requirement for the use of the improved $\chi^2$.
In this case it turns out that the p-value of the best
fit is low ($3.7\ 10^{-3}$), hence the $\Delta\chi^2$ test tends to be weaker than what is usually
expected. Thus we
have used here an approximate $\chi^2$ test instead, which explains why there is no 1 and 2 sigmas
contours. The result is shown in Figure~\ref{fig:fermiophobic}: it shows a four-fold degeneracy of the $\chi^2$ region with respect to the parameter space due to the 
obvious sign degeneracy of the two tree-level couplings. The black point 
corresponding to a fermiophobic Standard Model is excluded by more than 3.5 sigmas.

\subsection{Dilaton model}

Another interesting class of models is represented by dilatons, which can play the role of an impostor of the Higgs.
A dilaton can be thought of as a Pseudo-Nambu-Goldstone boson associated with an approximate scale invariance: it can give rise to phenomenology at 
colliders analogous to the one of a Higgs boson because it is expected to couple with the terms that break scale invariance, i.e. mass terms~\cite{dilaton1}.
Therefore, it can mimick a Higgs boson in Higgs-less models~\cite{dilaton2}, or modifying the couplings 
of a standard-like Higgs boson via mixing~\cite{dilaton3}.  
Dilatons are for example present in all extra-dimensional models, where they are associated with the compactification of the extra space dimensions, and in technicolour models, where they may appear as light scalar degrees of freedom of the confining theory~\cite{dilaton4}.
They have already been indicated as possible impostors of the Higgs.

\begin{figure}[htb]
\begin{center}
\includegraphics[width=0.45\textwidth]{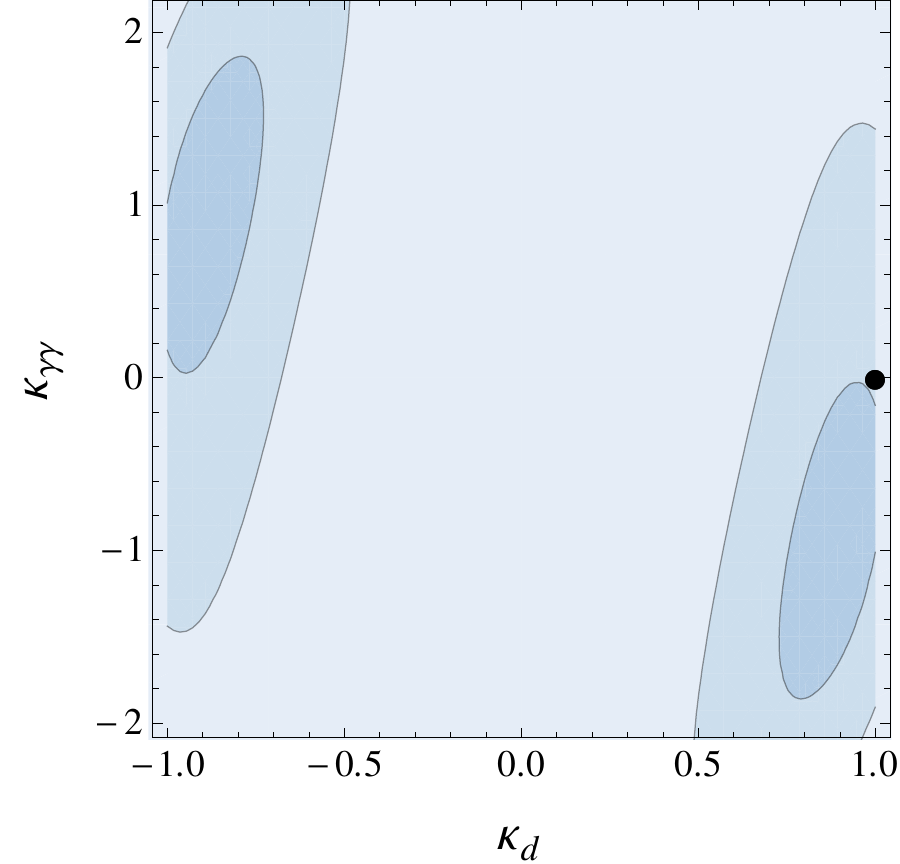}
\end{center}
\caption{\footnotesize Dilatonic model fit at the LHC for a Higgs boson with $m_H = 125$ GeV using
all channels from CMS.
Here we present the allowed region in a slice $\kappa_{gg} = 0$, in the $(\kappa_d, \kappa_{\gamma\gamma})$ plane. Darker (lighter) blue are 
the 1, 2 $\sigma$ regions.   }
\label{fig:dilaton}
\end{figure}

In this section, we will study a simplified dilaton model, where the impersonator has couplings to all massive states in the SM equal to the SM Higgs boson up to a rescaling factor $\kappa_d = v/f$, where $v$ is the SM Higgs vacuum expectation value, and $f$ is the scale associated with the breaking of the scale invariance (typically one expects $f > v$). 
The model under consideration, therefore, has equal tree level couplings $\kappa_W = \kappa_Z = \kappa_f = \kappa_d$.
Production cross sections and decay widths are accordingly 
modified, which allows us to get constraints on $\kappa_d$. We still study here in addition to this $\kappa_d$, the influence of new 
physics entering the loops, giving $\kappa_{gg}$ and $\kappa_{\gamma \gamma}$ coefficients, and thus perform a 3 parameter fit. 
We use here again the $\chi^2$ statistical test.
In Figure~\ref{fig:dilaton} we show a slice of the parameter space for $\kappa_{gg} = 0$, which includes the SM Higgs point ($\kappa_d = 1$, $\kappa_{\gamma\gamma}=0$).
The standard model like case  is consistent with the fit on the 
CMS data within a bit more than 1 sigma, as expected. The best fit corresponds to a slightly smaller values of $|\kappa_d| < 1$, however with new physics in the Higgs to 
$\gamma \gamma$ loop (when assuming $\kappa_{gg}=0$). 
In any case there is neither strong exclusion nor strong indication for such 
a dilation scenario in present data: the only information we can extract is that small values of $\kappa_d \ll 1$ are disfavoured, even by allowing for arbitrary new physics contributions in the loops.
If this conclusion were to hold a stronger statistical significance, any model of dilatons described by our simplified parameterisation would be excluded.

\section{Conclusion}

We have discussed a generalisation of the parameterisation proposed in \cite{Cacciapaglia:2009ky} to include tree-level couplings and
we showed how it can be used for testing and putting exclusion limits on models of new physics beyond the Standard Model. The most 
important radiative corrections, involving QCD NLO corrections, can be easily included in this parameterisation. 
We have compared this formalism to other parameterisations, in particular the 
one proposed in \cite{LHCHiggsCrossSectionWorkingGroup:2012nn}. 
We showed that the two, while sharing the same tree-level structure, are fundamentally different concerning the treatment the the 
loop-level couplings. In particular while the parameters in \cite{LHCHiggsCrossSectionWorkingGroup:2012nn} are inspired from the experimentally 
measured quantities, our parameterisation is tailored to investigate BSM models, keeping track of the specific correlations among the 
parameters. It also allows more easily to interpret mass limits and contributions to the loops giving the effective Higgs 
boson vertices. 
In fact, the parameters we propose are easily calculable in extensions of the Standard Model, and we clearly see in the plots that 
different models cluster in specific directions in the $\kappa_{gg}$ and $\kappa_{\gamma \gamma}$ parameter space.
This property is due to our choice to normalise the New Physics loop to the SM top one.
We also performed 2 parameter fits of the CMS and ATLAS results in the $H \to \gamma \gamma$, $H \to ZZ$ and 
$H \to b{\overline{b}}$ channels and 2 and 3 parameter fits using all available channels, showing that the 2 and 3 parameter fits 
already include all the necessary information and are therefore a good approximation at this stage. 
More precise measurements of extra channels will require the inclusion of more effective parameters.
We have also given few example of possible dedicated fits of BSM models: a  little Higgs model, a fermiophobic model and a dilation model, testing in all these cases the relevant parameters space with a $\chi^2$ test using available data.
We hope that this work will trigger interest of adding the proposed parameterisation to the existing ones in performing 
experimental fits on data by the ATLAS and CMS collaborations concerning the Higgs boson.

\section*{Acknowledgement}
We thank L. Panizzi for discussions and for suggesting the fit of the fermiophobic Higgs model.
We also thank S. Shotkin-Gascon, N. Chanon and the CMS group in Lyon for useful discussions.

\end{document}